\documentclass{article}

\usepackage{arxiv}

\usepackage[utf8]{inputenc} 
\usepackage[T1]{fontenc}    
\usepackage{hyperref}       
\usepackage{url}            
\usepackage{booktabs}       
\usepackage{amsfonts}       
\usepackage{nicefrac}       
\usepackage{microtype}      
\usepackage{lipsum}		
\usepackage{graphicx}
\usepackage{natbib}
\usepackage{doi}
\usepackage{amsmath}

\title{\emph{floodlight} - A high-level, data-driven sports analytics framework}

\author{ \href{https://orcid.org/0000-0001-7264-4575}{\includegraphics[scale=0.06]{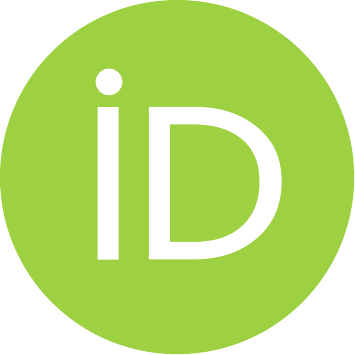}\hspace{1mm}Dominik~Raabe}\thanks{Corresponding Author} \\
	Institute of Exercise, Training and Sport Informatics\\
	German Sport University Cologne\\
	\texttt{dominik.raabe@dshs-koeln.de} \\
	\And
	\href{https://orcid.org/0000-0001-5660-9876}{\includegraphics[scale=0.06]{orcid.pdf}\hspace{1mm}Henrik~Biermann} \\
	Institute of Exercise, Training and Sport Informatics\\
	German Sport University Cologne\\
	\texttt{h.biermann@dshs-koeln.de} \\
	\And
	\href{https://orcid.org/0000-0002-9394-913X}{\includegraphics[scale=0.06]{orcid.pdf}\hspace{1mm}Manuel~Bassek} \\
	Institute of Exercise, Training and Sport Informatics\\
	German Sport University Cologne\\
	\texttt{m.bassek@dshs-koeln.de} \\
	\And
	\href{https://orcid.org/0000-0002-7403-7338}{\includegraphics[scale=0.06]{orcid.pdf}\hspace{1mm}Martin~Wohlan} \\
	Institute of Exercise, Training and Sport Informatics\\
	German Sport University Cologne\\
	\texttt{m.wohlan@dshs-koeln.de} \\
	\And
	\href{https://orcid.org/0000-0003-1300-9330}{\includegraphics[scale=0.06]{orcid.pdf}\hspace{1mm}Rumena~Komitova} \\
	Institute of Exercise, Training and Sport Informatics\\
	German Sport University Cologne\\
	\texttt{r.komitova@dshs-koeln.de} \\
	\And
	\href{https://orcid.org/0000-0001-7059-5319}{\includegraphics[scale=0.06]{orcid.pdf}\hspace{1mm}Robert~Rein} \\
	Institute of Exercise, Training and Sport Informatics\\
	German Sport University Cologne\\
	\texttt{r.rein@dshs-koeln.de} \\
	\And
	\href{https://orcid.org/0000-0003-1323-1628}{\includegraphics[scale=0.06]{orcid.pdf}\hspace{1mm}Tobias~Kuppens~Groot} \\
	Independent \\
	\texttt{tkuppensgroo@uni-osnabrueck.de} \\
	\And
	\href{https://orcid.org/0000-0002-3406-9175}{\includegraphics[scale=0.06]{orcid.pdf}\hspace{1mm}Daniel~Memmert} \\
	Institute of Exercise, Training and Sport Informatics\\
	German Sport University Cologne\\
	\texttt{d.memmert@dshs-koeln.de} \\
}

\renewcommand{\headeright}{Preprint}
\renewcommand{\undertitle}{}
\renewcommand{\shorttitle}{\textit{floodlight}}

\hypersetup{
pdftitle={floodlight - A high-level, data-driven sports analytics framework},
pdfsubject={cs.DS},
pdfauthor={Dominik~Raabe, Henrik~Biermann, Manuel~Bassek, Martin~Wohlan, Rumena~Komitova, Tobias~Kuppens~Groot, Robert~Rein, Daniel~Memmert},
pdfkeywords={},
}

\begin{document}
\maketitle

\begin{abstract}
	The present work introduces \textit{floodlight}, an open source Python package built to support and automate team sport data analysis. It is specifically designed for the scientific analysis of spatiotemporal tracking data, event data, and game codes in disciplines such as match and performance analysis, exercise physiology, training science, and collective movement behavior analysis. It is completely provider- and sports-independent and includes a high-level interface suitable for programming beginners.	The package includes routines for most aspects of the data analysis process, including dedicated data classes, file parsing functionality, public dataset APIs, pre-processing routines, common data models and several standard analysis algorithms previously used in the literature, as well as basic visualization functionality. The package is intended to make team sport data analysis more accessible to sport scientists, foster collaborations between sport and computer scientists, and strengthen the community's culture of open science and inclusion of previous works in future works.
\end{abstract}


\section{Summary}

The increase of available data has had a positive impact on the entire sports domain and especially sport science \citep{Morgulev2018}. Two major data sources of relevance in this domain are spatiotemporal tracking data of athlete positions as well as manually annotated match event data \citep{Stein2017, Memmert2018}. These two data types are regularly collected by professional sport organizations in different team invasion games such as football, basketball, or handball \citep{Memmert2021}. These data sources open up a whole range of new analysis possibilities across multiple (sub)disciplines in the field, including match and performance analysis, exercise physiology, training science, or collective movement behavior analysis. As an example, player tracking data has been used extensively to analyze physical \citep{Castellano2014} as well as tactical \citep{Rein2016} performance in football.

The \textit{floodlight} Python package provides a framework to support and automate team sport data analysis. \textit{floodlight} is constructed to process spatiotemporal tracking data, event data, and other game meta-information to support scientific performance analyses. \textit{floodlight} was designed to provide a general yet flexible approach to performance analysis, while simultaneously providing a user-friendly high-level interface for users with basic programming skills. The package includes routines for most aspects of the data analysis process, including dedicated data classes, file parsing functionality, public dataset APIs, pre-processing routines, common data models and several standard analysis algorithms previously used in the literature, as well as basic visualization functionality.

\begin{figure}
	\centering
	\includegraphics[scale=.4]{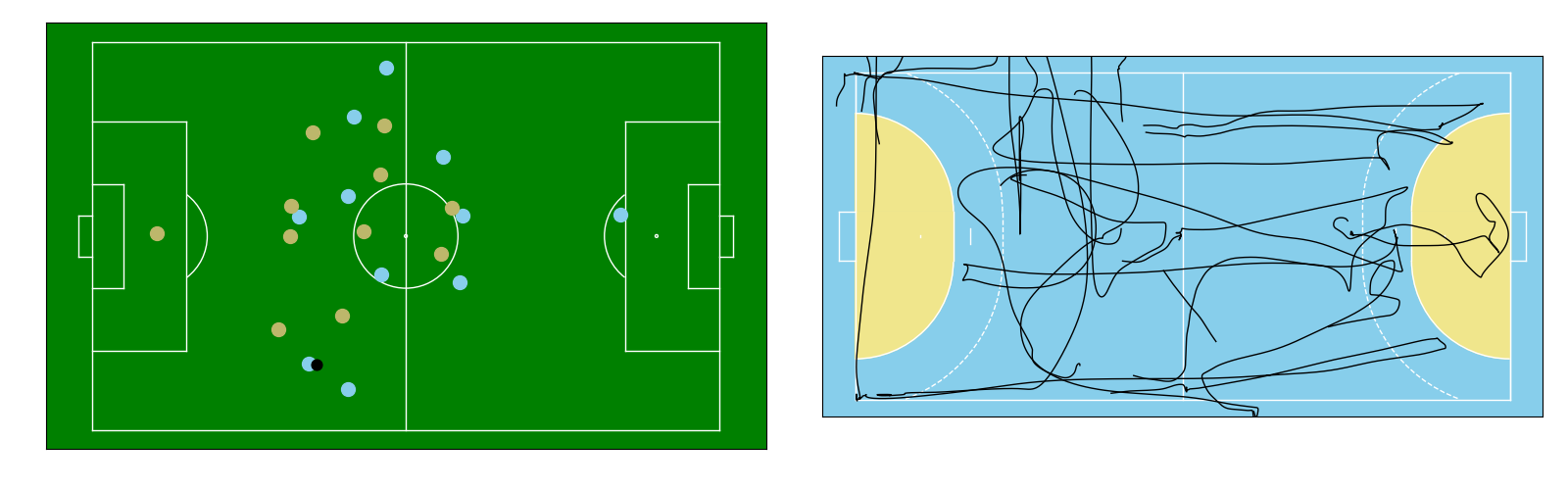}
	\caption{Positions of football players (left) and trajectories of handball players (right) from real-world match data as visualized with \textit{floodlight}.}
	\label{fig:sample}
\end{figure}

Central to the package is a set of generalized, provider- and sports-independent core data structures based on \textit{numpy} \citep{Harris2020} and \textit{pandas} \citep{McKinney2010}. Each of these data structures are dedicated to one specific type of sports data, including spatiotemporal tracking data, event data, game codes (meta information such as ball possession information), pitch information regarding the embedding of data and playing surfaces into Cartesian coordinate systems, as well as team and player properties (such as frame-wise velocity or acceleration values). The data structures are designed with a focus on scientific computing, i.e., optimized for accessible and intuitive data manipulations as well as sensitive to performance by utilizing \textit{numpy}'s view-, vectorization- and indexing techniques.

The core data classes allow internal storage and processing of sports data whilst decoupling from any format-specific requirements. Consequently, \textit{floodlight} is built around these objects, comprising several elementary modules of the data processing pipeline. For data loading, the package provides parsing submodules with functions that dissect and map data from specific provider formats to core data structures (including providers such as Kinexon, Tracab, Stats Perform, StatsBomb, or DFL), which eliminates problems caused by the many, strongly varying data formats in use. Data loaders and mappers for available public datasets such as the EIGD-H dataset \citep{Biermann2021} are additionally included. In terms of data processing, the package provides dedicated manipulation functionality such as spatial transformations helpful for spatial data synchronization or signal filters based on \textit{scipy} \citep{Virtanen2020}. For data inspection, basic visualization functionality based on the \textit{matplotlib} package \citep{Hunter2007} is included (see Figure \ref{fig:sample}).

The actual data analysis part is realized by a submodule providing several data models. These models provide a toolbox of domain-specific data analysis procedures from different subdomains such as exercise physiology, e.g., the metabolic power model \citep{diPrampero2018}, dynamical system approaches, e.g., approximate entropy \citep{Pincus1991}, or collective tactical behavior, e.g., centroid-based measures \citep{Sampaio2012, Bourbousson2010}. All models follow the same syntax inspired by the \textit{scikit-learn} package \citep{Buitinck2013}, where upon instantiation, a central fitting method is called with core data structures. Subsequently, required computations can be queried with additional class methods. This allows a consistent syntax and collection of similar measures into cohesive data models while limiting the repetition of basic calculations and allowing simple future extensions.

\section{Statement of Need}

Despite the increase in volume, the technical requirements for team sport data analysis have constantly remained high. This can be partially attributed to the complexity and heterogeneity of the data itself \citep{Stein2017, Memmert2018}, but also to multiple practical and theoretical challenges. These include the necessity of complex file parsing procedures for provider-specific data formats, low compatibility across data providers, or differing standards for spatial or temporal resolution of data, often requiring specialized pre-processing routines. Meeting these challenges typically requires massive and customized overhead programming in sports data analysis projects. At the same time, there hardly exist any general, proprietary or open source, software alternatives which can be used out of the box for scientific purposes. Existing software is either commercially driven (i.e., proprietary, limited to a specific data provider or focused on industrial applications), or task-specific (i.e., limited to a certain data source, data format, sport or subtask) which leaves the problem of adapting code to multiple different APIs within the analysis process.

These current constraints resulted in a situation where a typical analysis workflow requires the (re)implementation of each processing pipeline module in its entirety with respect to the specific project's needs. For sport scientists who typically lack programming skills (which are usually not part of their formal training) this can become an insurmountable hurdle. As a consequence, advanced team sports data analyses remain inaccessible for large parts of the sport scientific community which poses a significant hindrance for future progress. Accordingly, the \textit{floodlight} package was designed to specifically address this problem and significantly ease advanced analyses of sports data. \textit{floodlight} automates standard data processing routines and provides a high-level interface accessible to users with just basic programming skills. The \textit{floodlight} documentation contains several tutorials as well as an extensive compendium discussing the technical aspects of team sports data analysis to ensure easy access and understanding of the routines and their design choices. The tutorials increase the beginner-friendliness of \textit{floodlight} and allow its usage in educational settings, e.g., for team sport data analytics courses.

Another hurdle faced by sports scientist relates to the current lack of collaboration and code sharing practices within the field. At present, sharing proposed data models or algorithms for analyses is the exception rather than the rule. In parts, the lack of sharing often stems from the proprietary nature of the raw data, but is further exacerbated by lack of data format gold standards. More generally, disciplines that employ team sports data analysis have reported a culture that contains very little replications and works incorporating previous findings \citep{Herold2019}, low applicability of research by practitioners \citep{Bishop2008, Mackenzie2013, Herold2019} and limited interdisciplinary approaches between computer and sport scientists \citep{Rein2016, Goes2021}. A major milestone in the process of meeting these challenges is to find feasible ways of sharing data and algorithms \citep{Rein2016}. The \textit{floodlight} package can be seen as a first step in this direction with a toolbox-approach collecting common data manipulation and processing techniques.

\textit{floodlight} will therefore be equally useful for sports scientists as well as computer scientists, working in academia or applied settings. The package will therefore serve to bring these users groups together and foster future interdisciplinary collaborations. Ideally, this will also promote further open source contributions that share advanced data processing algorithms in the domain and enable future work incorporating previous findings.

\section{Example}

The following code sample illustrates how \textit{floodlight} reduces a typical performance analysis pipeline to just a few lines of code. In the example, one sample of data is queried from the public EIGD-H dataset, filtered, and the cumulative metabolic work of the home team is calculated for the entire segment of data:

\begin{center}
    \begin{verbatim}
    from floodlight.io.datasets import EIGDDataset
    from floodlight.transforms.filter import butterworth_lowpass
    from floodlight.models.kinetics import MetabolicPowerModel
    
    dataset = EIGDDataset()
    home_team_data, away_team_data, ball_data = dataset.get()
    
    home_team_data = butterworth_lowpass(home_team_data)
    
    model = MetabolicPowerModel()
    model.fit(home_team_data)
    metabolic_power = model.cumulative_metabolic_power()
    \end{verbatim}
\end{center}

\section*{Acknowledgements}

This project has received funding from the German Federal Ministry
of Education and Research (BMBF) to the last author under grant number 01IS20021A.

\section*{Resources}

More information on the project as well as source code and documenation can be found here:

Package hosting: \url{https://pypi.org/project/floodlight/}

Project source code: \url{https://github.com/floodlight-sports/floodlight}

Documentation: \url{https://floodlight.readthedocs.io}

\bibliographystyle{unsrtnat}
\bibliography{main}	

\end{document}